\documentstyle[12pt,epsf,epsfig,wrapfig]{article}
\begin{document}
\begin{center}
{\large \bf
Semi-inclusive spin asymmetries in polarised deep inelastic electron
scattering 
\\ }
\vspace{5mm}
\underline{E.E.W. Bruins} for the HERMES Collaboration
\\
\vspace{5mm}
{\small\it
Laboratory for Nuclear Science, Massachusetts Institute of Technology, \\
77 Massachusetts Avenue, Cambridge, MA 02139-4307 \\
Temporary Address: DESY, Notkestrasse 85, 22607 Hamburg, Germany
\\ }
\end{center}
\begin{center}
ABSTRACT

\vspace{5mm}
\begin{minipage}{130 mm}
\small
In 1995, the HERMES experiment at DESY (Hamburg, Germany) has 
taken its first data, using the 27.5 GeV polarised  
HERA positron beam impinging on a longitudinally polarised 
internal $^3$He target. Preliminary results from the semi-inclusive
physics program are presented. 
\end{minipage}
\end{center}
Polarised deep-inelastic lepton-nucleon scattering has provided 
information about the spin structure of the nucleon, which 
can be expressed in terms of spin distribution functions of
partons in the infinite momentum frame. 

\begin{wrapfigure}{r}{8cm}
\epsfig{figure=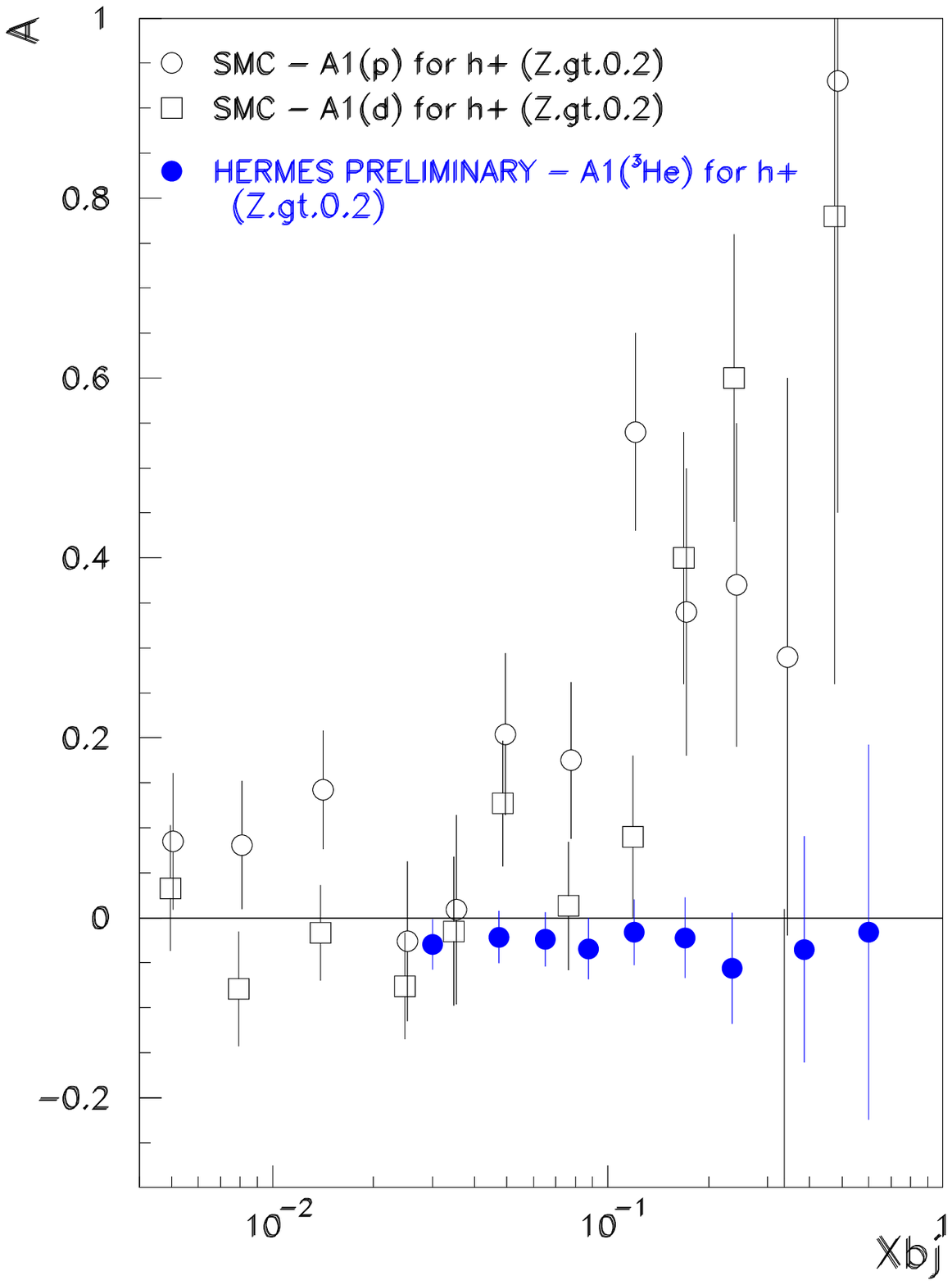,width=8cm}
{\small Figure 1: Spin asymmetry of semi-inclusive, positively charged 
hadrons, for the polarised H, D (SMC) and $^3$He (HERMES,
preliminary)}
\end{wrapfigure}
Up to now, most information has come from inclusive measurements, 
detecting the scattered lepton only. It is expected
that the investigation of the hadronic final state in the scattering 
process will allow a deeper insight into the relevant processes than 
the inclusive measurement alone.
In HERMES [1], the scattered positron and the hadronic final state are
measured in a wide kinematic range and with a good angular 
and momentum resolution. Positrons have been clearly identified 
and separated from the hadrons using preshower, calorimeter and 
\v Cerenkov counter information. Semi-inclusive pions
were identified and separated from heavier hadrons by means of the 
\v Cerenkov counter (above 5.6 GeV/c) and time-of-flight techniques 
(below 2 GeV/c).
Semi-inclusive final state hadrons which decay in the detector 
volume have been identified by the reconstruction of the invariant
mass of the decaying system. A full description of the experiment,
the detector set up and the target can be found in [2].

The analysis of the 1995 semi-inclusive data is in progress. We focus 
on both the polarised and unpolarised physics potential of HERMES,
and expect to deliver the first complete quark spin decomposition on 
the polarised neutron.
In Figure~1 the $^3$He spin asymmetry is shown for positively 
charged semi-inclusive hadrons $A_1^{h^+}$($^3$He) 
for $z = p/\nu > 0.2$, where $p$ is the momentum of the hadron and $\nu$
the energy transfer in the reaction. 
The data indicate a small negative asymmetry, in agreement with the naive
expectation based on charge/isospin conjugation and existing SMC data
for polarised $H$ and $D$, measured on polarised butanol and deuterated
butanol targets [3]. 
Monte Carlo studies with PEPSI [4], based
on the LUND fragmentation model, have shown that, due to the forward 
acceptance of the HERMES detector, an even lower
$z$-cut does not significantly change the underlying physics in our data.
In Figure~2 we show the spin asymmetry for semi-inclusive neutral pions, 
identified by the
reconstruction of the invariant mass of the two decay gammas.
\begin{wrapfigure}{r}{8cm}
\epsfig{figure=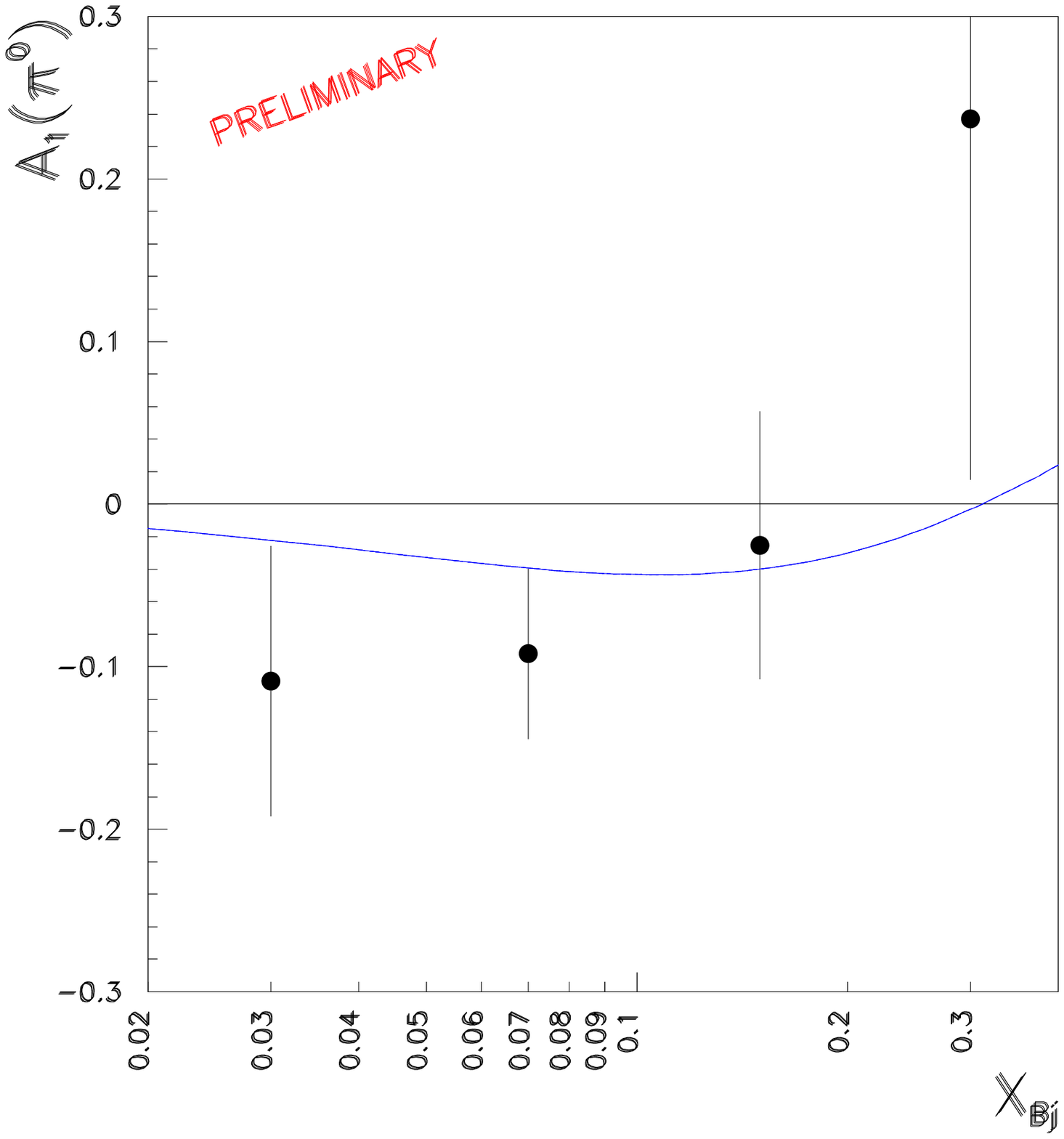,width=8cm}
{\small Figure 2: Spin asymmetry of semi-inclusive, neutral pions
for polarised $^3$He (HERMES, preliminary). The curve shows a 
parton model prediction using the Sch\"afer parametrisation [5].}
\end{wrapfigure}

Due to the relatively high \v Cerenkov pion threshold (5.6 GeV/c) during
the 1995 run, the spin asymmetry for {\it identified} charged pions has large 
statistical uncertainties. 
The 1996 data on the proton will profit from a lower threshold,
and lead to a significant improvement in our semi-inclusive pion data. 
Using linear combinations of the various semi-inclusive spin 
asymmetries and (model dependent) Monte Carlo predictions, we thus expect to 
extract quark spin distributions as a function of $x_{Bj}$ 
from the $^3$He and proton data, separately.
Finally, the combination of the 1995 and 1996 data will allow
an extraction of the valence quark spin distributions $\Delta u_v$ and $\Delta d_v$ as
a function of $x_{Bj}$.\\ 
\\
Preliminary results for unpolarised $\rho$ production have been extracted.
These indicate the superior quality of the data which HERMES is expected
to deliver in the 
forthcoming years. Figure~3 shows the decay angle distribution of
elastic, diffractive $\rho$ mesons, without subtraction
of the (small) non-exclusive and combinatorial background. Figure~4 shows
the nuclear transparency of $^3$He (normalized to $H+D$ data from less than 24 
hours dedicated data taking) to the incoherently produced sub-sample of
the above mentioned $\rho$ mesons. Although the statistics is 
low and the systematics not yet fully understood, combined with our
1996 proton data, we expect to measure the nuclear transparency of
$^3$He as a function of momentum and energy transfer in the reaction.
Targets with $A>3$ are being considered at the moment.
\begin{wrapfigure}{r}{15cm}
\epsfig{figure=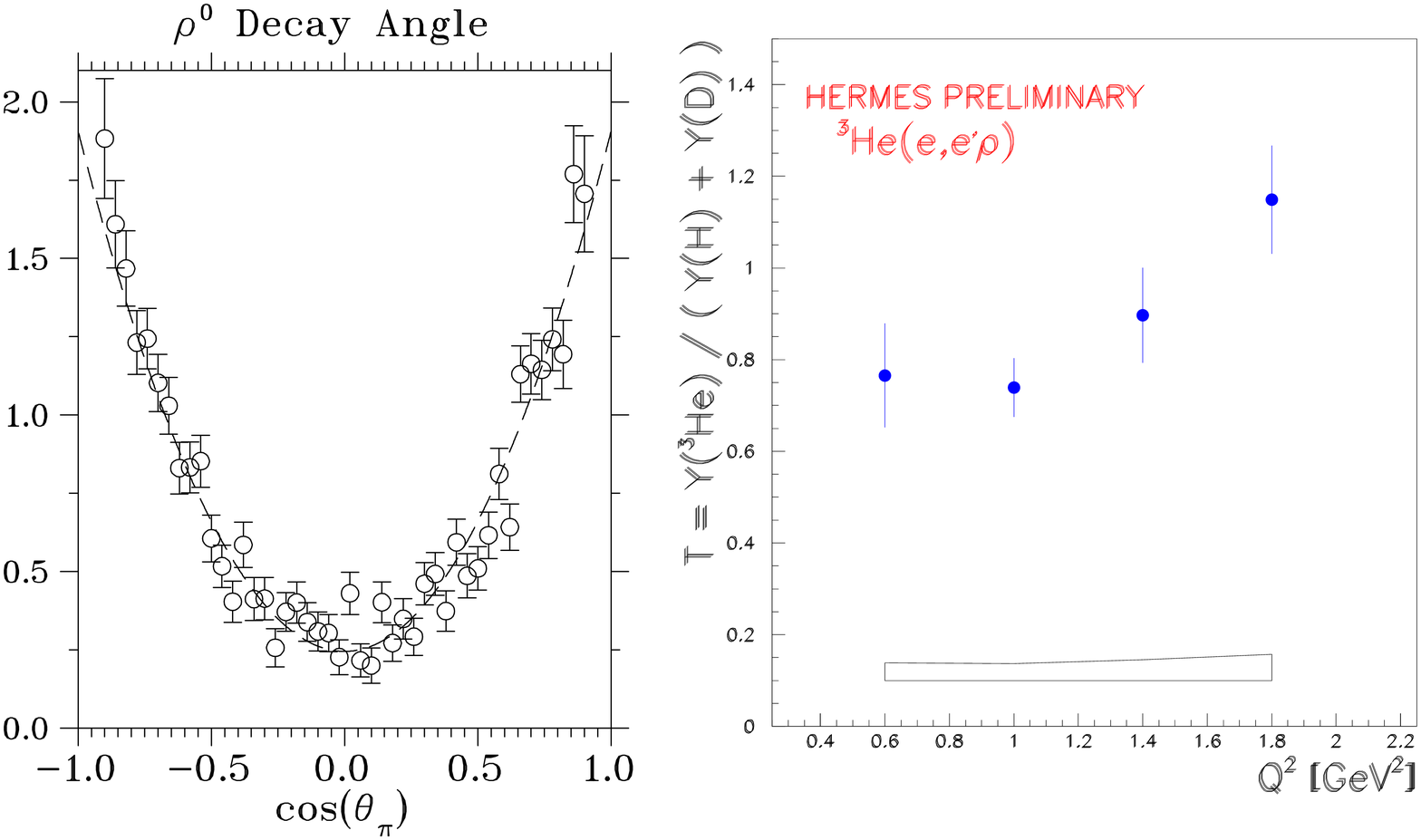,width=15cm}
{\small Figure 3: (a) The decay angle distribution of elastic diffractive
rho mesons from $^3$He (No background subtraction);
(b) Nuclear transparency of $^3$He to incoherently produced
elastic, diffractive rho mesons, as a function of $Q^2$. 
We allow for a preliminary 20 \% overall
normalisation uncertainty (HERMES, preliminary)}
\end{wrapfigure}

\vspace{2.cm}
{\small\begin{description}
\item{[1]} 
K.~Coulter {\it et al.}, DESY-PRC 90/01 (1990);
M.~D\"uren, DESY-HERMES 95-02 (1995)
\item{[2]} A.~Simon, this symposium
\item{[3]} B.~Adeva~{\it et al.}, SMC Collaboration, Phys. Rev. {\bf B369} (1996) 93
\item{[4]}
G.~Ingelmann, LEPTO -- CERN programming pool W5045 (1986);
M.~Veltri, H.~Ihssen, PEPSI -- HERMES Collaboration (1995)
\item{[5]} A.~Sch\"afer {\it et al.}, hep-ph/9505306
\end{description}}
\vfill

\end{document}